\documentclass[apjl]{emulateapj}

\usepackage{graphicx}

\newcommand{\fnl}{f_\mathrm{NL}}

\shortauthors{G.\ De Troia {\it et al.}}

\begin{document}
  
  
\title{Searching for non Gaussian signals in the BOOMERanG 2003 CMB maps}
  

\author{
G.~De~Troia\altaffilmark{1,2},  
P.A.R.~Ade\altaffilmark{4}, 
J.J.~Bock\altaffilmark{5,15},
J.R.~Bond\altaffilmark{6},
J.~Borrill\altaffilmark{7,17},
A.~Boscaleri\altaffilmark{8},
P.~Cabella\altaffilmark{21},
C.R.~Contaldi\altaffilmark{6,16},
B.P.~Crill\altaffilmark{9},
P.~de~Bernardis\altaffilmark{2},
G.~De~Gasperis\altaffilmark{1},
A.~de~Oliveira-Costa\altaffilmark{13},
G.~Di~Stefano\altaffilmark{12},
P.~G.~Ferreira\altaffilmark{21},
E.~Hivon\altaffilmark{11},
A.H.~Jaffe\altaffilmark{16},
T.S.~Kisner\altaffilmark{7,17},
M.~Kunz\altaffilmark{22},
W.C.~Jones\altaffilmark{5,15},
A.E.~Lange\altaffilmark{15},
M.Liguori\altaffilmark{24},
S.~Masi\altaffilmark{2},
S.~Matarrese\altaffilmark{23},
P.D.~Mauskopf\altaffilmark{4},
C.J.~MacTavish\altaffilmark{6},
A.~Melchiorri\altaffilmark{2,18},
T.E.~Montroy\altaffilmark{28},
P.~Natoli\altaffilmark{1,19},
C.B.~Netterfield\altaffilmark{14},
E.~Pascale\altaffilmark{14},
F.~Piacentini\altaffilmark{2,25},
D.~Pogosyan\altaffilmark{20},
G.~Polenta\altaffilmark{2},
S.~Prunet\altaffilmark{11},
S.~Ricciardi\altaffilmark{2,26},
G.~Romeo\altaffilmark{12},
J.E.~Ruhl\altaffilmark{3},
P. Santini\altaffilmark{2},
M.~Tegmark\altaffilmark{13},
M. Veneziani\altaffilmark{2,27}, and
N.~Vittorio\altaffilmark{1,19}.
}

\altaffiltext{1}{ Dip. Fisica, Universit\`a Tor
Vergata, Roma, Italy}
\altaffiltext{2}{ Dip. Fisica, Universit\`a La
Sapienza, Roma, Italy} 
\altaffiltext{3}{ Physics Department, Case Western Reserve University,
		Cleveland, OH, USA}
\altaffiltext{4}{ Dept. of Physics and Astronomy, Cardiff University, 
		Cardiff CF24 3YB, Wales, UK} 
\altaffiltext{5}{ Jet Propulsion Laboratory, Pasadena, CA, USA}
\altaffiltext{6}{Canadian Institute for Theoretical Astrophysics, 
		University of Toronto, Toronto, Ontario, Canada}
\altaffiltext{7}{ Computational Research Division, Lawrence Berkeley National Laboratory, Berkeley, CA, USA}
\altaffiltext{8}{ IFAC-CNR, Firenze, Italy}
\altaffiltext{9}{ IPAC, California Institute of Technology, Pasadena, CA, USA}
\altaffiltext{11}{ Institut d'Astrophysique, Paris, France}
\altaffiltext{12}{ Istituto Nazionale di Geofisicae Vulcanologia, Roma,~Italy}
\altaffiltext{13}{ Dept. of Physics, Massachusetts Institute of Technology, Cambridge,  MA, USA}
\altaffiltext{14}{ Physics Department, University of Toronto, Toronto, Ontario, Canada}
\altaffiltext{15} {Observational Cosmology, California Institute of
Technology, Pasadena, CA, USA}
\altaffiltext{16} {Theoretical Physics Group, Imperial College, London, UK}
\altaffiltext{17} {Space Sciences Laboratory, UC Berkeley, CA, USA}
\altaffiltext{18} {INFN, Sezione di Roma 1, Roma, Italy}
\altaffiltext{19}{ INFN, Sezione di Tor Vergata, Roma, Italy}
\altaffiltext{20} {Dept.\ of Physics,University of Alberta,Edmonton,AB,Canada}
\altaffiltext{21} {Astrophysics, University of Oxford, Keble Road, Oxford OX1 3RH, UK}
\altaffiltext{22} {D\'epartement de Physique Th\'eorique, Universit\'e de Gen\`eve, Switzerland}
\altaffiltext{23} {Dipartimento di Fisica G. Galilei, Universit\`a di Padova and INFN, Sezione di Padova, Italy}
\altaffiltext{24} {Department of Applied Mathematics and Theoretical Physics, University of Cambridge, UK}
\altaffiltext{25} {European Space Astronomy Centre (ESAC), European Space Agency, Madrid, Spain}
\altaffiltext{26} {INAF- Osservatorio Astronomico di Padova,Italy}
\altaffiltext{27} {APC, 10 rue Alice Domon et Lonie Duquet,75,Paris Cedex 13}
\altaffiltext{28} {Sierra Lobo, Inc. 11401 Hoover Rd. Milan, OH 44846, USA}



\begin{abstract}
We analyze the BOOMERanG 2003 (B03) 145 GHz temperature map
  to constrain the amplitude of a non Gaussian, primordial
  contribution to CMB fluctuations. We perform a pixel space analysis
  restricted to a portion of the map chosen in view of high
  sensitivity, very low foreground contamination and tight control of
  systematic effects. We set up an estimator based on the three
  Minkowski functionals which relies on high quality simulated data,
  including non Gaussian CMB maps. We find good agreement with the
  Gaussian hypothesis and derive the first limits based on BOOMERanG
  data for the non linear coupling parameter $\fnl$ as $-300<\fnl<650$
  at $68\%$ CL and $-800<\fnl<1050$ at $95\%$ CL.
\end{abstract}

\keywords{cosmology: cosmic microwave background}

\section{Introduction}
While cosmology is entering its precision era, the target of
experiments aimed at the Cosmic Microwave Background (CMB) is shifting
towards weak signals, e.g.\ polarization, the SZ effect and non
Gaussian (NG) features. Detection of NG signals can be of significant
help in constraining the mechanisms that explain the generation of
cosmological perturbations. Provided that systematic effects will not
degrade high sensitivity CMB mapping, present and future experiments
could in principle be sensitive to non linearities due to second order
effects in perturbation theory~\cite{bartolo}. This signal is usually
parametrized by a non linear coupling factor $\fnl$ that controls the
level of a quadratic contribution to the primordial gravitational
potential $\Phi$ \cite{komatsu_spergel}:
\begin{equation}\label{eqn:fnl}
\Phi(\mathbf{x})=\Phi_G(\mathbf{x})+\fnl \left[\Phi_G(\mathbf{x})^2 -
\left\langle \Phi_G(\mathbf{x})^2 \right\rangle \right]
\end{equation}
where $\Phi_G$ is a zero mean, Gaussian random field.

Several groups have reported NG constraints on CMB data. All
suborbital efforts to date have found no significant deviation from
Gaussianity in the CMB field: MAXIMA-1 reported $|\fnl|<950$ at
$1\sigma$~\cite{santos,wu}, while VSA found an upper limit of 5400 at
$2\sigma$ \cite{vsa}; Archeops has recently improved their limits to
$-800<\fnl<1100$ ($2\sigma$), although their analysis is based on
assumptions only valid for the large angular scales dominated by the
Sachs-Wolfe effect~\cite{archeops}. The BOOMERanG 1998 dataset has also
been tested for Gaussianity, both in pixel \cite{polenta} and in
Fourier \cite{detroia} space, finding no trace of NG signals.
However, BOOMERanG has set no $\fnl$ limit so far. One of the purposes
of this paper is to provide such limits with the analysis of the new
2003 data. The limits presented here are more stringent than those
found by any suborbital experiment to date, properly accounting for
sub-horizon angular scales.

The WMAP team constrained $\fnl$ to be $-54 < \fnl < 114$ 
\cite{spergel}. Using an improved version of the WMAP team estimator Creminelli et
al.\ (2007) set the most stringent limits to date at $-36 < \fnl <
100$. Thus the $\fnl$ analysis does not show any departure from
Gaussianity in WMAP data. However, some authors have looked at general
deviations from Gaussianity (i.e. not based on any specific
parametrization of NG) and claimed highly-significant detection of
NG features in the WMAP maps \cite{copi,vielva,cruz}.

In this paper we perform a pixel space analysis of the B03
temperature (T) data set, using the observed field's moments and Minkowski
functionals (MFs) to build Gaussianity tests. We assess the statistical
significance of our results comparing the data to a set of highly
realistic, Gaussian Monte Carlo (MC) simulated maps. In order to
constrain $\fnl$, we build a goodness of fit statistics based on MFs
and calibrated against a set of NG CMB maps, that are generated
according to the algorithm set forth in \cite{liguori}. 

The plan of this letter is as follows: in
section~\ref{sec1} we briefly describe the B03 experiment,
the dataset it has produced and our simulation pipeline. In
section~\ref{sec2} we compute the map's moments and MFs
of the data and compare results against Gaussian MC
simulated maps. Then we derive constraints for $\fnl$.
Finally, in section~\ref{sec4} we draw our main conclusions.

\section{The BOOMERanG 2003 dataset}
\label{sec1}
The balloon borne B03 experiment has been flown from Antarctica
in 2003. It gathered data for 14 days in
three frequency bands, centered at 145, 245 and 345 GHz. Here
we restrict ourselves to the 145 GHz data that are most sensitive to
CMB fluctuations. These have been obtained with polarization
sensitive bolometers (PSB). The analysis of the dataset has produced high quality maps 
of the southern sky that have been conveniently divided in
three regions: a ``deep'' (in terms of integration time)
survey region ($\sim 90$ square degrees) and a ``shallow'' survey
region ($\sim 750$ square degrees), both at high Galactic latitudes,
as well as a region of $\sim 300$ square degrees across the Galactic
plane. The deep region is completely embedded in the shallow
region. Here we only consider a subset of the data that
contains all of the deep region and part of the shallow, for a total
of 693 square degrees (1.7\% of the sky). The mask
we use is square, 26 degrees in side, centered at
about RA=$82.6^\circ$ and DEC=$-44.2^\circ$, and excludes all detected
point sources in the field. 
This region has been selected in view of high sensitivity 
CMB observation with low foreground contamination and was observed 
with a highly connected
scanning strategy to keep systematics under control. We use the
T data map reduced jointly from eight PSB at 145 GHz~\cite{masi}. 
In this region, the signal rms on $3.4'$
pixels is $\sim 90~\mu\mathrm{K}$ and instrumental noise has an rms
of $\sim 20~\mu\mathrm{K}$ in the deep region and of
$\sim 90~\mu\mathrm{K}$ in the shallow region. In
harmonic space, binned estimates of the CMB angular power spectrum
retain signal to noise $>1$ well beyond $\ell \sim 1000$. One may
compare these figures with WMAP: in the three year release, WMAP
combined sensitivity in the region observed by B03 is $\sim
100~\mu\mathrm{K}$ on $3.4'$ pixels, close to WMAP's mean
pixel error. However, WMAP's beams are broader than B03, 
so its $\ell$ space error is
$\sim$ 5 times larger than B03 at $\ell\simeq 1000$. On the other
hand, B03 has not been devised to measure multipoles at $\ell\lesssim
50$. In this sense, our NG analysis probes angular scales
complementary to those constrained by WMAP.
While we do not consider here the Stokes Q and U polarization maps,
our T map has been marginalized with respect to
linear polarization. For a description of the
instrument and the measured T and polarization maps
see~\cite{masi} and for the CMB TT, TE and EE power spectra
see~\cite{jones,piacentini,montroy}.

To assess the robustness of our tests of Gaussianity we use a set of
simulated MC maps that mimic the B03 data. To produce these,
we follow the same steps performed when analysing real data. The
Gaussian CMB sky signal is simulated from the power
spectra that best fits the B03 data
\cite{mactavish}. 
This signal is smoothed according to the measured beam
and synthetized into a pixelized sky map, using Healpix
routines \cite{healpix}.  Using the B03 scanning strategy, the
signal map is projected onto 8 timestreams, one for each 145 GHz
detector.  Noise only timestreams are also produced, as Gaussian realizations of each
detector's noise power spectral density, estimated from
the data accounting for cross talks among detectors. The timelines
are reduced with the ROMA mapmaking code~\cite{natoli,degasperis} replicating
the actual flight pointing and transient flagging, to produce
T,Q and U maps. With this procedure, we
can simulate signal, noise and signal plus noise timestream.

To constrain $\fnl$ we use MC simulations of 
NG CMB maps obtained from a primordial 
gravitational potential of the form given in eq. (\ref{eqn:fnl}).
These maps have been produced including first order CMB 
radiative transfer effects \cite{liguori}. The power spectrum of the NG
maps is identical to that of the Gaussian CMB
simulations.

\section{Tests of Gaussianity and constraints on $\fnl$}
\label{sec2}
\begin{figure}
\begin{center}
\includegraphics[width=4cm,height=3cm]{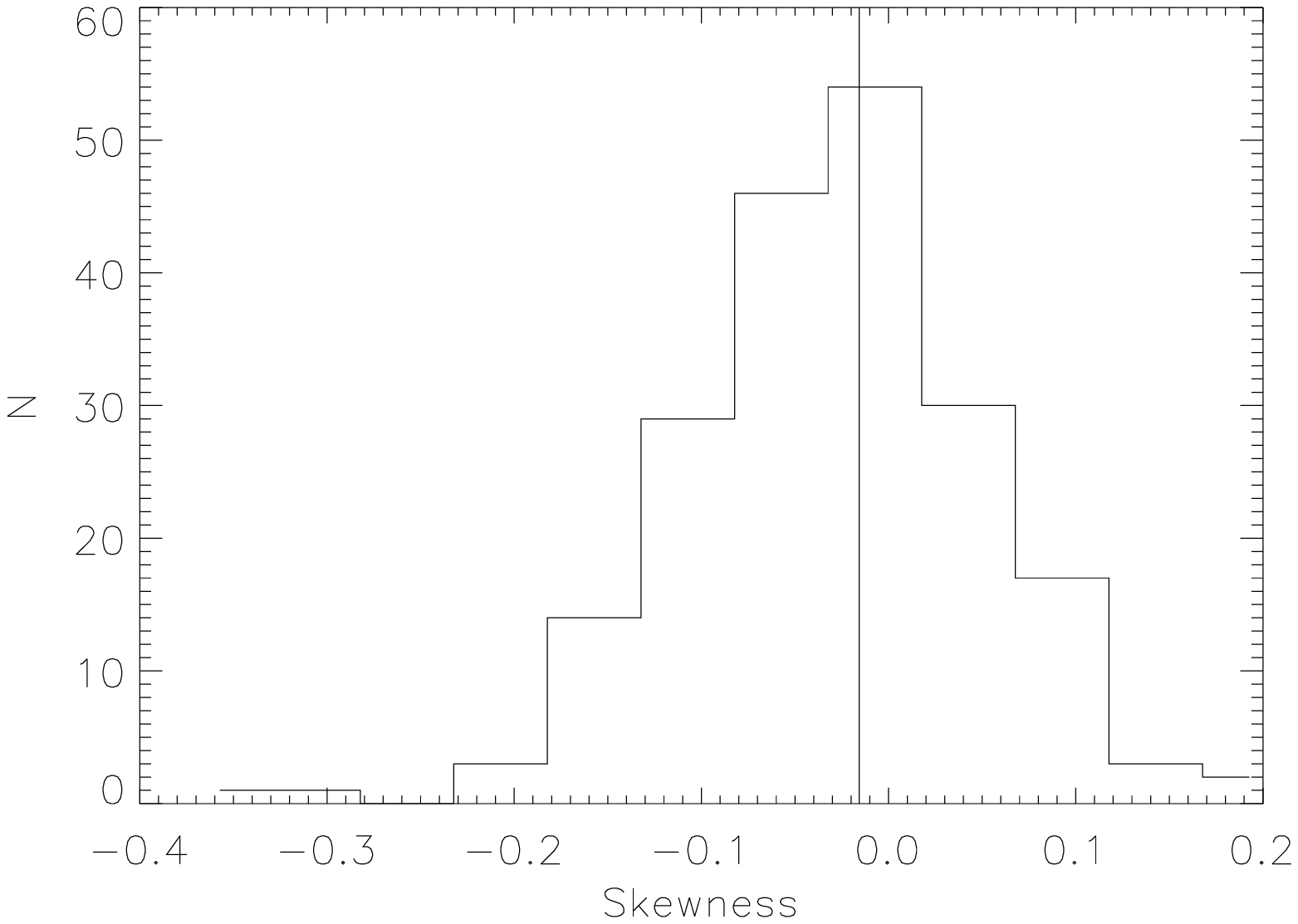}
\includegraphics[width=4cm,height=3cm]{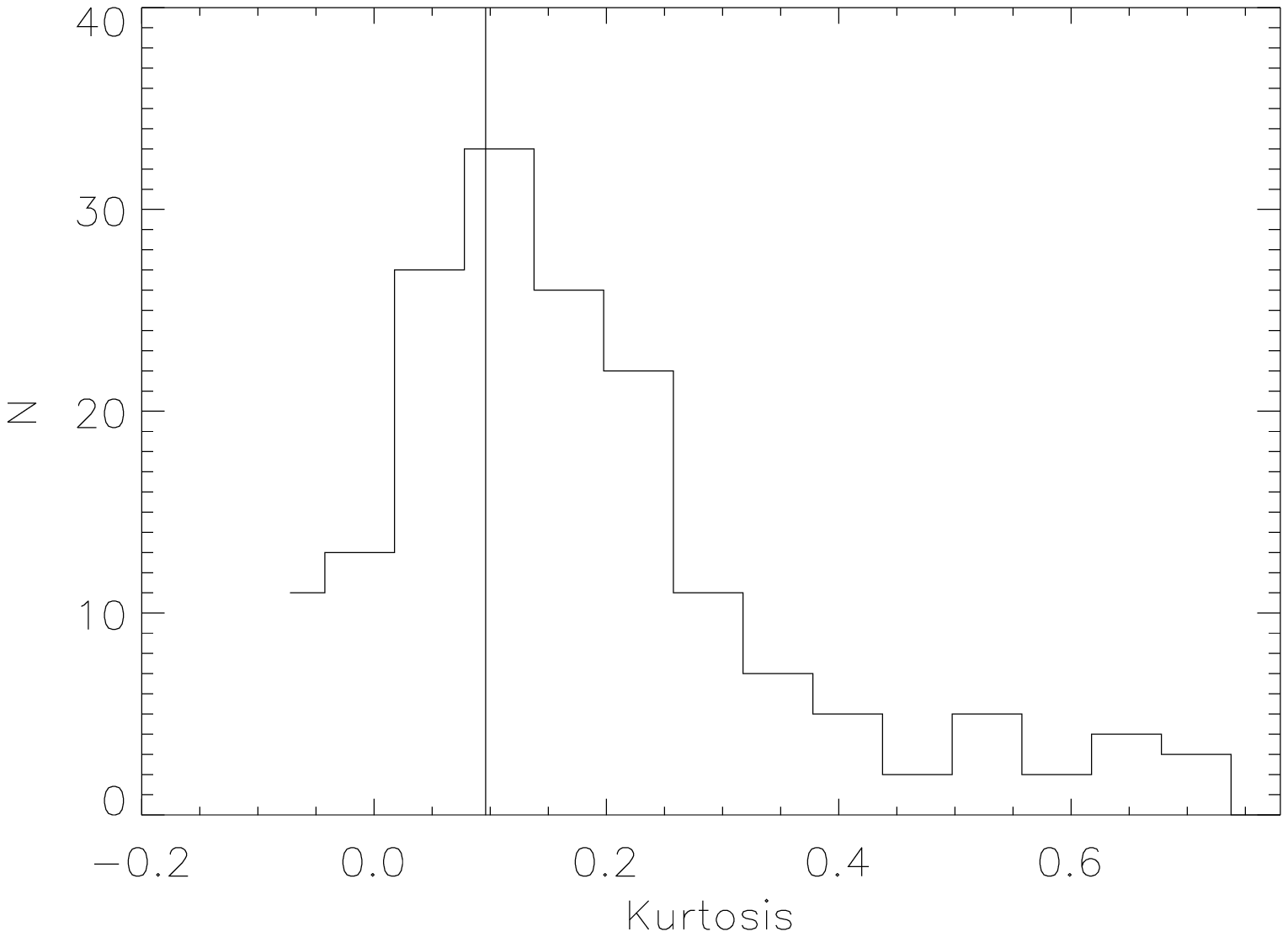}
\caption{The left panel shows the distribution of $S_3$ 
estimator calculated from the
200 Gaussian MC maps. The $S_3$ 
value of the B03 data is represented by the vertical line. 
The right panel shows the same for $S_4$} 
\label{figura1}
\end{center}
\end{figure}

Working at $6.8'$ Healpix resolution ($N_\mathrm{side}=512$), 
we first compute the normalised skewness
$S_3$ and kurtosis $S_4$ of our pixelized field $T_i$. These are
obtained from the variance $\sigma^2=1/(N-1)\sum_i(T_i-\langle
T\rangle)^2$ and from the third and the fourth moment
$\mu_3=\sum_i(T_i-\langle T\rangle)^3/N$ and $\mu_4=\sum_i(T_i-\langle
T\rangle)^4/N$, where $N$ is the total number of pixels of the map and
$\langle T\rangle = \sum_i T_i/N$ its mean. We have $S_3=\mu_3/\sigma^3$, $S_4=\mu_4/\sigma^4-3$. From the data we get $S_3=-0.016$ and $S_4=0.096$. 
These values are plotted in Fig.~\ref{figura1} as a vertical 
line and compared to the
empirical distribution as derived from the MC
(signal plus noise)  maps. From the latter we
compute the probability $P(S_3^{\mathrm{sim}} >
S_3^\mathrm{data})=58\%$ and $P(S_4^\mathrm{sim} >
S_4^\mathrm{data})=77\%$. Hence, for these tests the data are
compatible with the Gaussian hypothesis.  The same tests are repeated
after having degraded the map to $13.6'$, finding similar
results.

\begin{figure}
\begin{center}
\includegraphics[width=8cm,height=7.5cm]{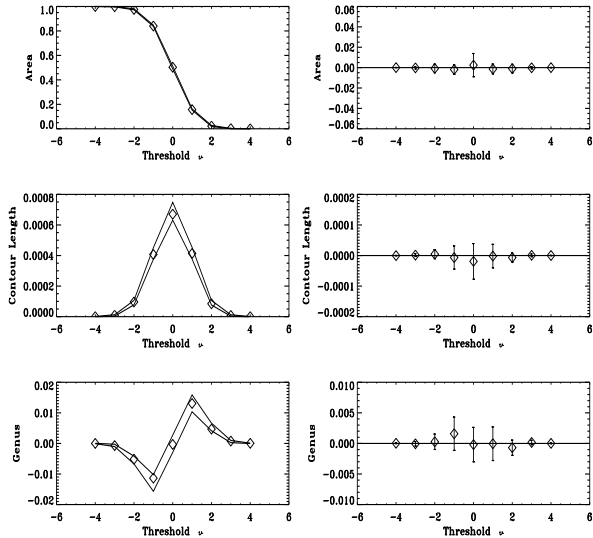}
\caption{ Left panels show the MFs estimated from  
the B03 data (diamonds) and the 2$\sigma$ confidence limits from the 200  
Gaussian MC maps. From top to bottom, the results correspond to the  
MFs M0, M1 and M2, respectively. Right panels show the residuals between the MC mean  
and the data.}
\label{figura2}
\end{center}
\end{figure}

To analyse the map with MFs~\cite{mink_f} we
consider the excursion sets $Q$ defined as the map's subsets exceeding a given
threshold $\nu$: $Q(\nu)=\{T_i:(T_i-\langle T\rangle)/\sigma >
\nu\}$. The three
MFs measure the total area
of the surviving regions of the map ($M_0$), their total contour
length ($M_1$), and the genus of the distribution which is related to
the difference between the number of ``hot'' and ``cold'' regions
($M_2$). For a Gaussian field the expectation values of the
functionals depend on a single parameter $\tau$:
$\langle
M_0(\nu)\rangle=\frac{1}{2}\left[1-\mathrm{erf}\left(\frac{\nu}{\sqrt{2}}\right)\right]$,
$\langle
M_1(\nu)\rangle=\frac{\sqrt{\tau}}{8}\mathrm{exp}\left(-\frac{\nu^2}{2}\right)$,
$\langle M_2(\nu)\rangle=\frac{\tau}{\sqrt{8\pi^3}}\nu
\mathrm{exp}\left(-\frac{\nu^2}{2}\right)$. In the case of a pure CMB
signal (no noise), $\tau$ is given by
$\tau=\frac{1}{2}\sum_{\ell=1}^{\infty}(2\ell+1)\,\ell(\ell+1)C_{\ell}$
\cite{schmalzing,winitzki}. Hence, $M_1$ and $M_2$ depend on the power
spectrum $C_\ell$. It is hence critical that the simulations
reproduce the model $C_\ell$'s that best fits the data. We work in
the flat sky limit, projecting our T values on the plane
locally tangent to the map \cite{cabella}. In Fig.~\ref{figura2} we 
plot MFs for the
data and $2\sigma$ limits set by 200 Gaussian simulations, as well as
the data residuals and their (again, $2\sigma$) errors. 

The functionals are computed at $9$ thresholds evenly spaced between $-4\sigma$ and
$+4\sigma$. 

\begin{figure}
\begin{center}
\includegraphics[width=4cm,height=2.5cm]{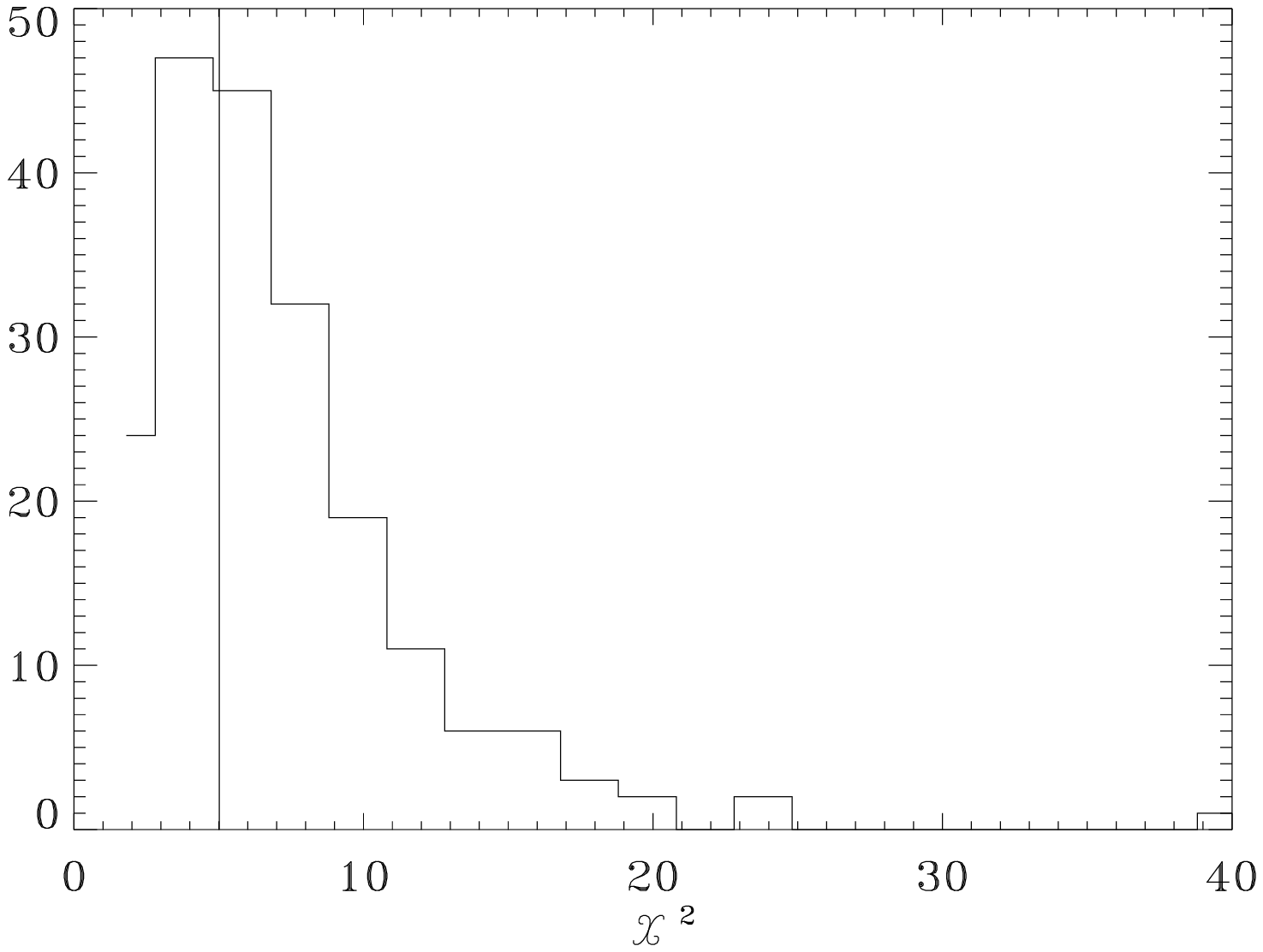}
\includegraphics[width=4cm,height=2.5cm]{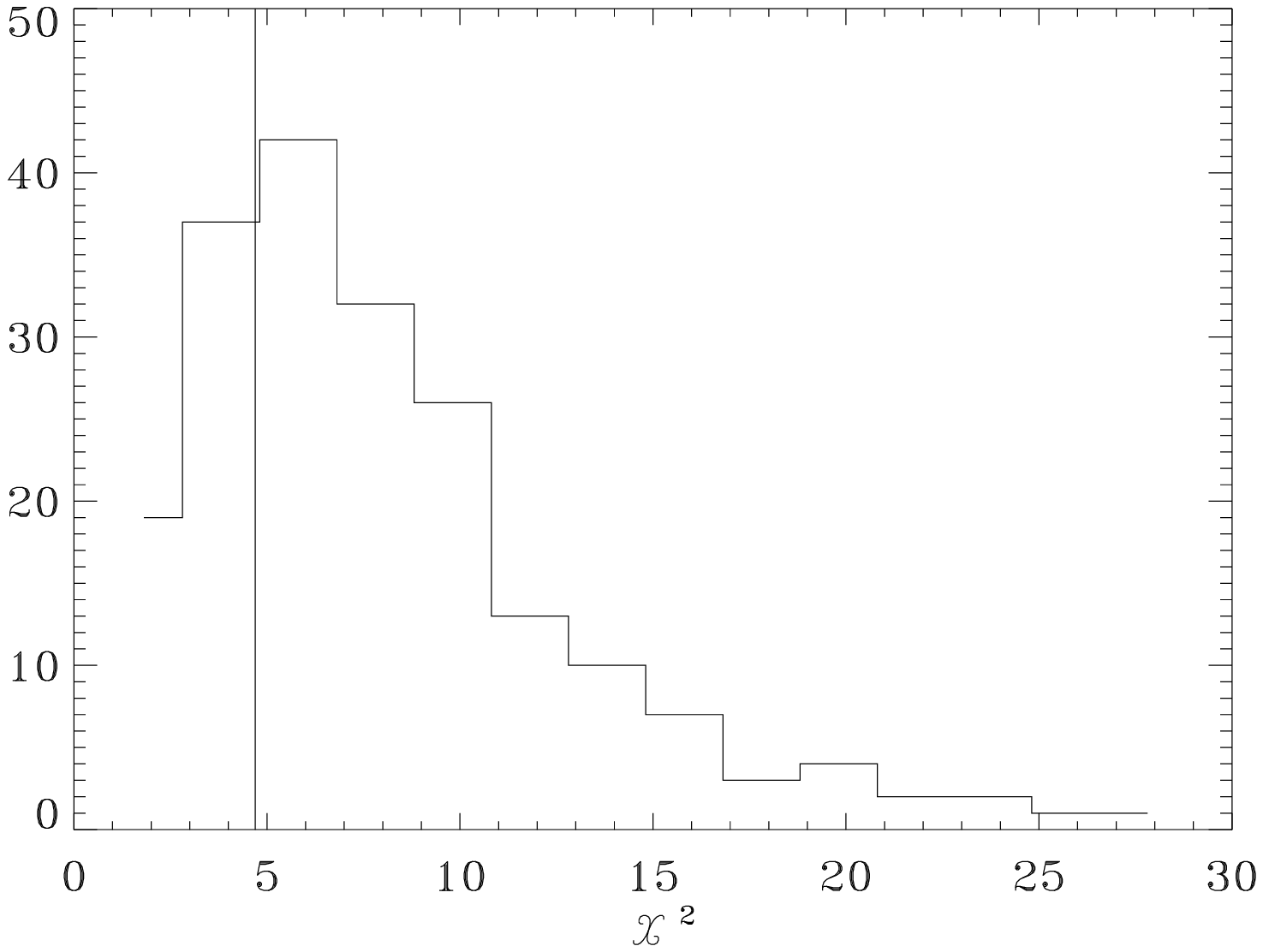}
\includegraphics[width=4cm,height=2.5cm]{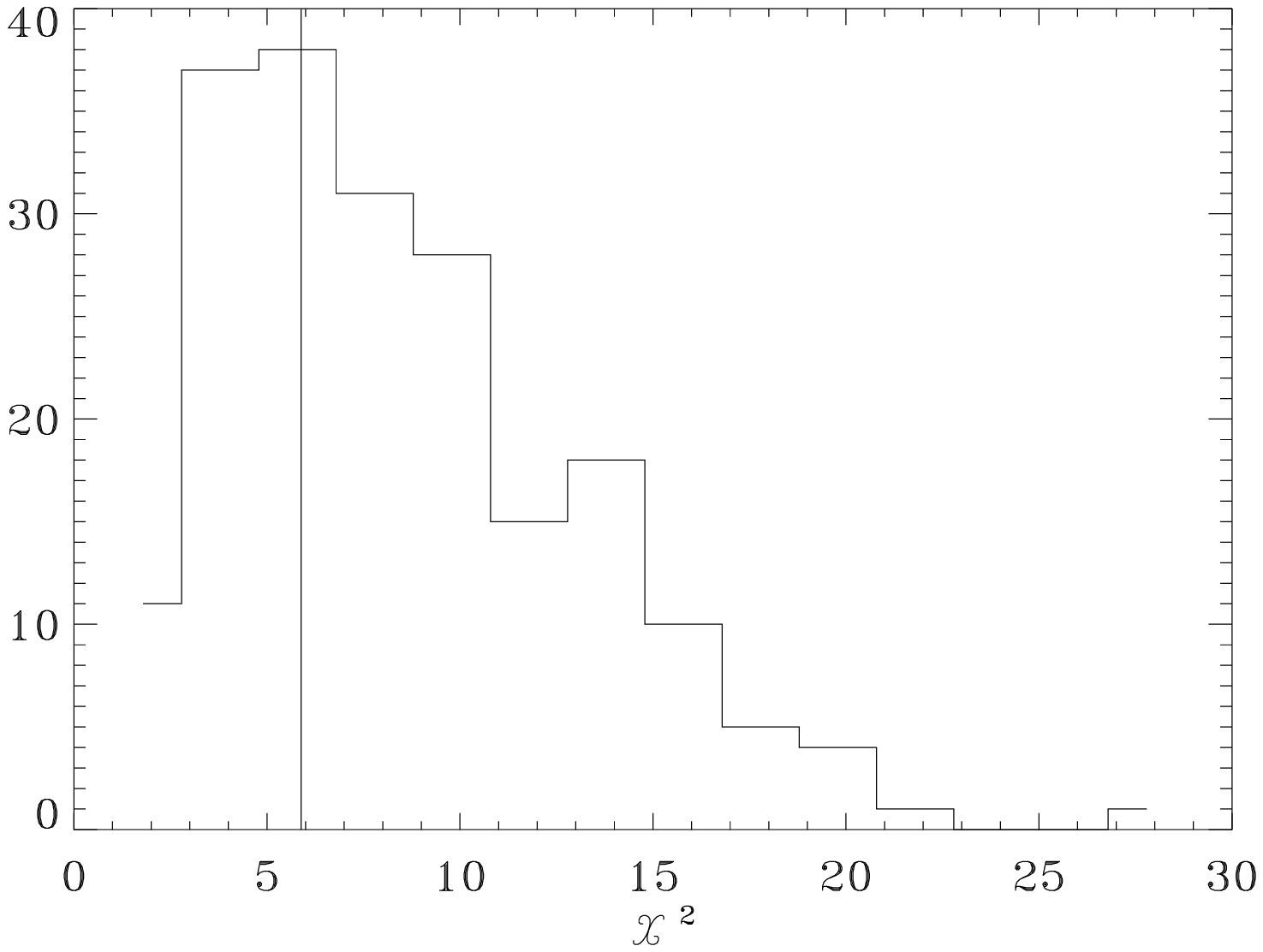}
\includegraphics[width=4cm,height=2.5cm]{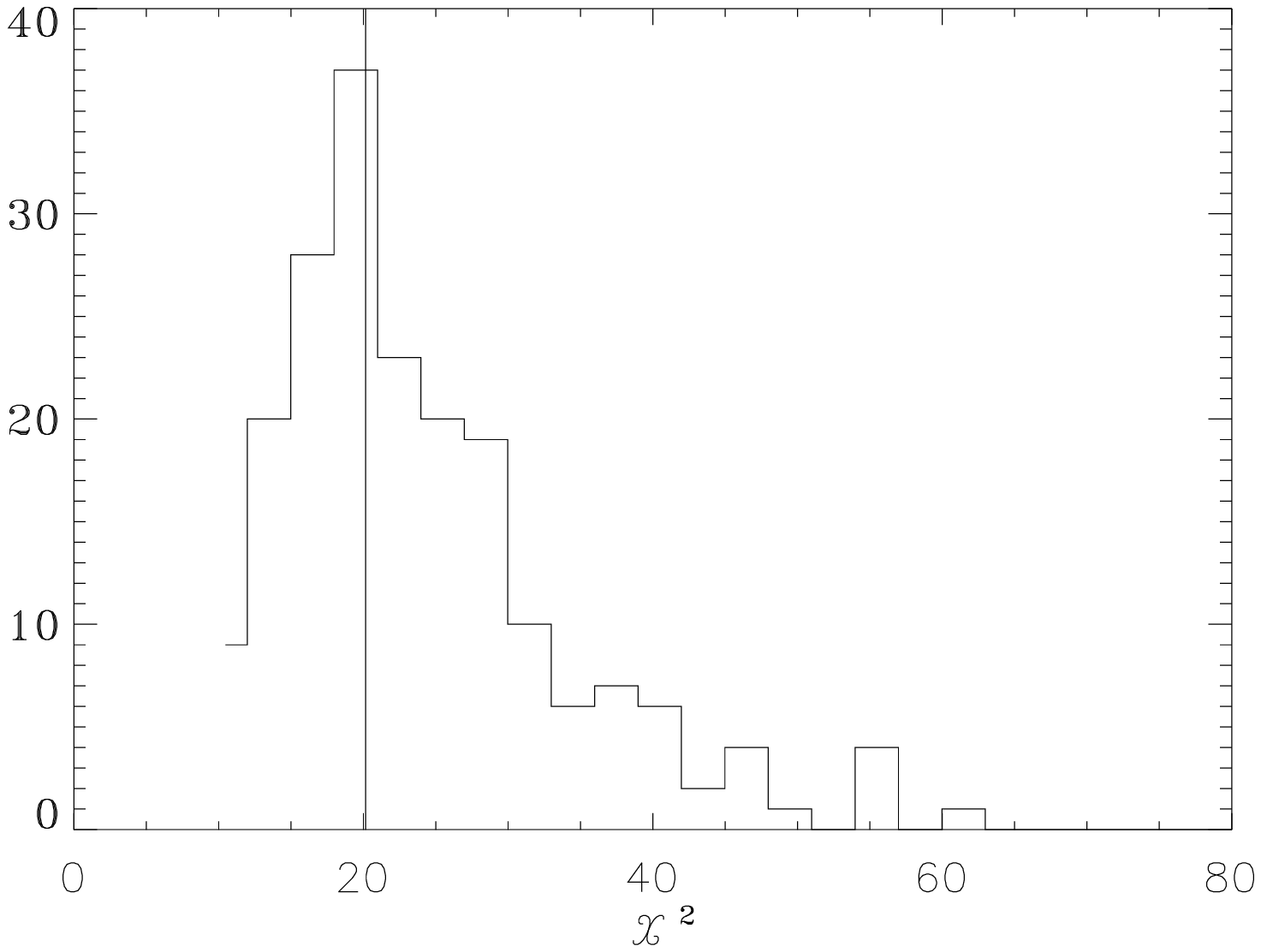}
\caption{The $\chi^2_B$ distribution (histogram) of MC simulated maps
and data value (vertical line) for the MFs. Top: 
area and contour length. Bottom: genus and ``joint'' estimator (see text).}
\label{figura3}
\end{center}
\end{figure}

Using the MC maps we can define a $\chi^2$ statistic:
\begin{equation}\label{chisqr}
\chi^2_{B,i}=\sum_{\nu\nu^{\prime}}(M_i^B-\langle M^{sim}_i\rangle)_{\nu}C^{-1}_{i,\nu\nu^{\prime}}(M_i^B-\langle M^{sim}_i\rangle)_{{\nu}^{\prime}}.
\end{equation}
Here $M_i^B$ ($ M^{sim}_i$) is any of the three MFs obtained from the
data (simulations), $\langle \cdot \rangle$ is the mean value
over MC realizations, and $C_{i,\nu\nu^{\prime}}=\langle
(M_{i,\nu}-\langle M_{i,\nu}\rangle) (M_{i,\nu^{\prime}}-\langle
M_{i,\nu^{\prime}}\rangle)\rangle$ is a covariance matrix, estimated
from an independent set of $\sim 1000$ Gaussian maps. In the first
three panels of Fig.~\ref{figura3} we show $\chi^2_i$ for each MF
(vertical line), plotted along with the empirical distribution sampled
via MC. We can define a ``joint'' estimator by grouping the $M_i$'s in
a single, $27$ elements data vector $M_J \equiv
\{M_0,M_1,M_2\}$ and defining a $\chi^2_B$ as a trivial extension of
Eq.~\ref{chisqr}. It is important that the covariance matrix built
for the joint estimator correctly accounts for correlations among
different functionals. However, we have found that in order to pin
down to a stable level these off-block couplings, one requires a
number realizations significantly higher than the $\sim 1000$ used
throughout our analysis. The latter number cannot be realistically
increased to desired level, because the GLS map making procedure is a
demanding computational task, even for the 
supercomputers we have used. Fortunately, we have found that using
white (instead of correlated) noise to estimate the
covariance matrix has a negligible impact on the analysis. This
finding can indeed be justified a posteriori, because the GLS map
making procedure is very effective in suppressing noise
correlations, that 
contribute very weakly to the estimator's final covariance.
The joint $\chi^2$ of the data is displayed as
the fourth panel in Fig.~\ref{figura3}, along with the MC empirical
distibution. The probability $P(\chi^2 > \chi^2_B)$ that a Gaussian map has a
larger $\chi^2$ than the B03 map is $76\%$ for $M_0$, $83\%$ for
$M_1$, $76\%$ for $M_2$ and $67\%$ for the ``joint'' estimator. The
values are fully consistent with the Gaussian hypothesis. We conclude
that our pixel space analysis does not detect any sign of NG behavior
in the B03 data.

\begin{figure}
\begin{center}
\includegraphics[width=4cm,height=3.5cm]{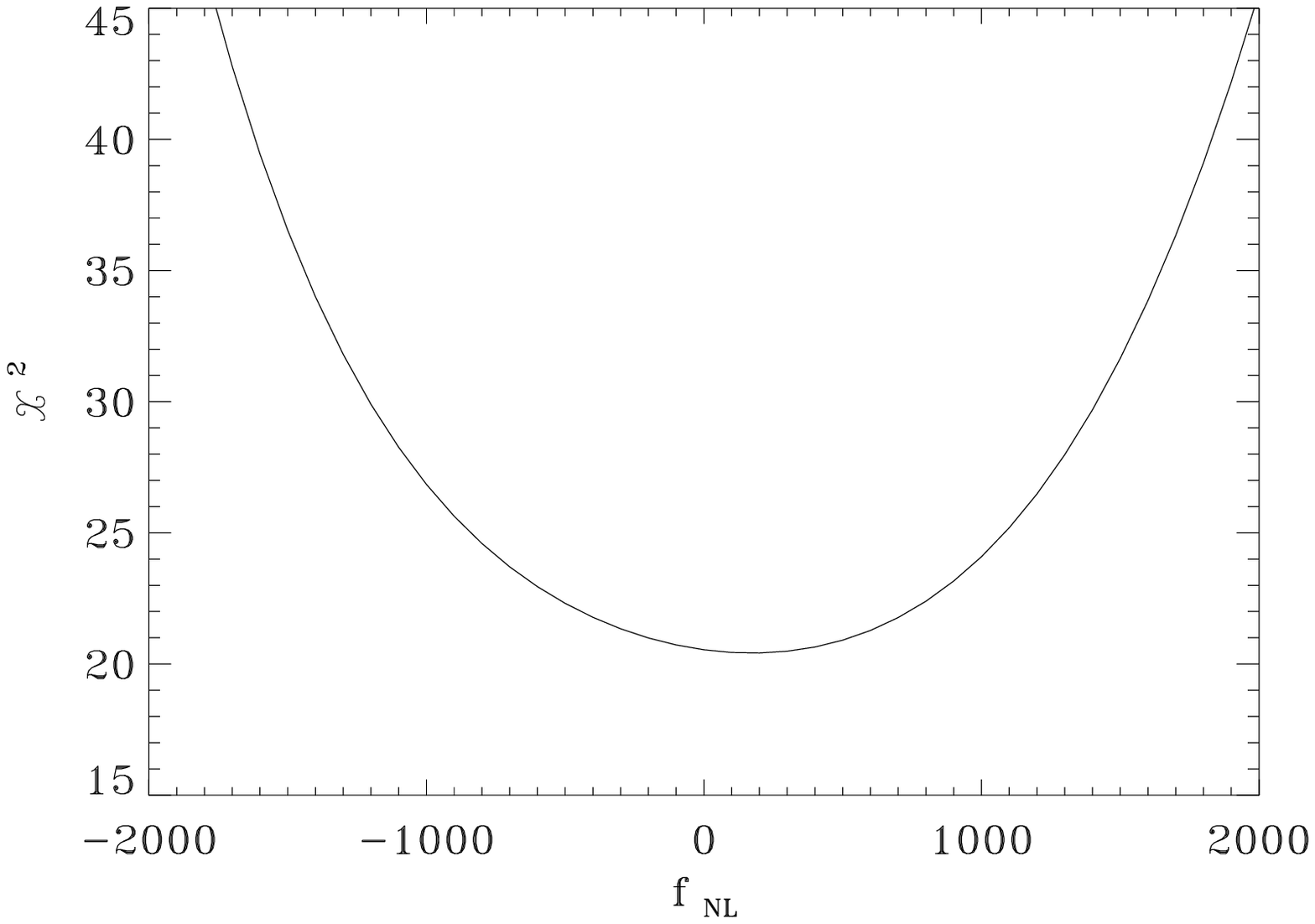}
\includegraphics[width=4cm,height=3.5cm]{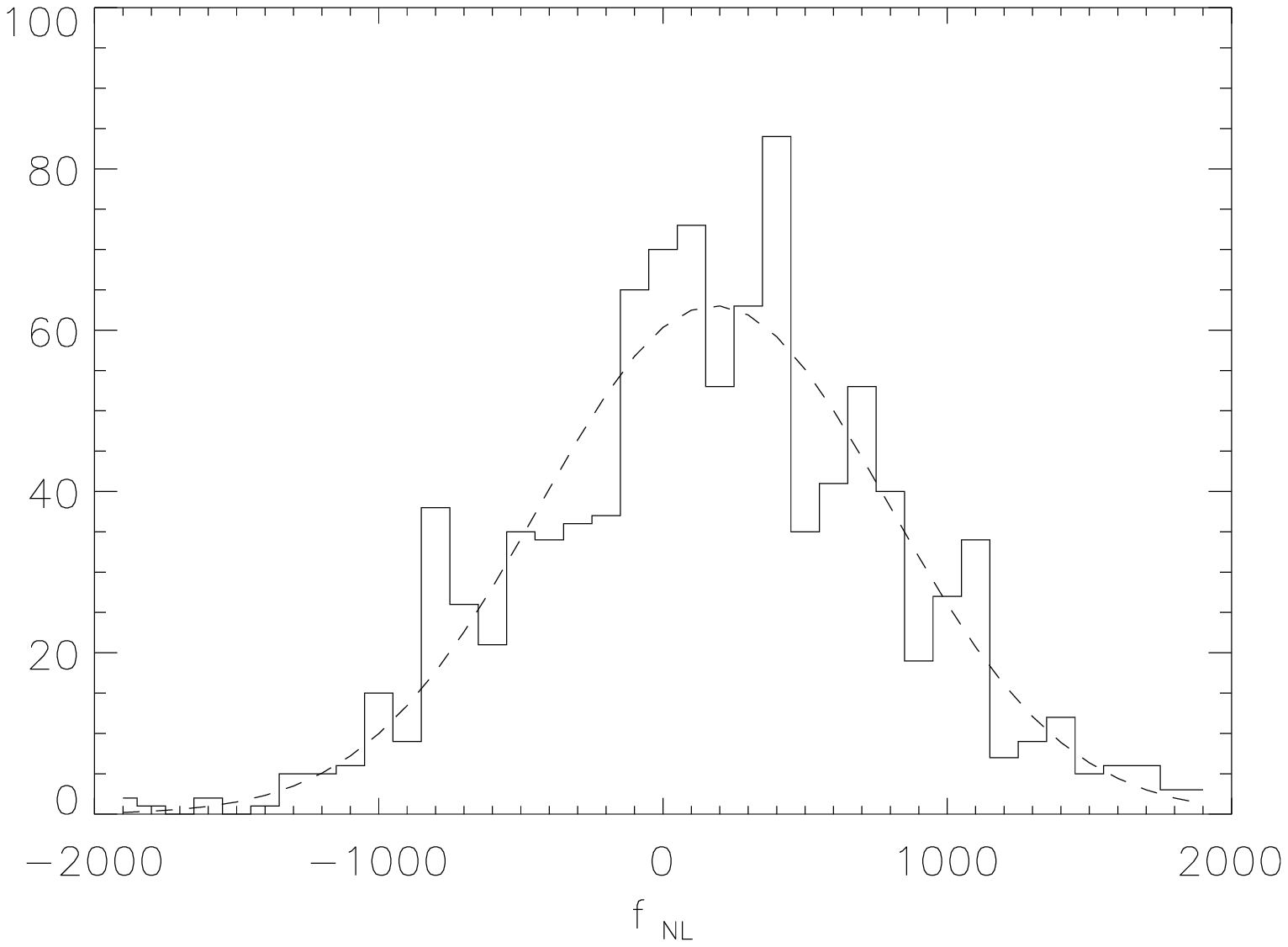}
\caption{The left panel shows the B03 data $\chi^2$ ($\chi^2_B$ in the text) as a function of $\fnl$, while the panel on the right shows the empirical $\chi^2$ distribution for NG maps ($\fnl=200$), sampled using 1000 simulations}
\label{figura4}
\end{center}
\end{figure}

We now want to constrain the quadratic coupling
parameter $\fnl$ defined in Eq.~\ref{eqn:fnl}.  
To this purpose we simulate NG CMB realizations
in the following way: firstly, we generate 
the Gaussian and NG part of the 
primordial potential defined by Eq.~(\ref{eqn:fnl});
then we convolve them with CMB first order radiation transfer 
functions to get the final CMB sky.
In this way we produce 200
$\mathbf{G}$ (Gaussian) maps and 200 $\mathbf{NG}$
counterparts (each $\mathbf{G}$ map has a uniquely defined
$\mathbf{NG}$ counterpart), so that for a given $\fnl$ our (signal only)
map is $\mathbf{G} + \fnl *\mathbf{NG}$. By adding noise maps, we
can define MF estimators in the spirit of the
section \ref{sec2}, with the difference that they are now functions of $\fnl$:
$J_B(\fnl)=M_J^B-\langle M_J(\fnl)\rangle$ (we only consider the
``joint'' estimator in what follows). Consequently, we now define the
data $\chi^2$ as $\chi^2_B(\fnl)=J_B(\fnl)^T \mathbf{C}^{-1}J_B(\fnl)$.
While in principle the covariance of the $M_J$'s is a function of
$\fnl$, this dependence is expected to be weak and is usually
neglected \cite{komatsu}. We have tested for this explicitely by using
our NG simulations. We plot $\chi^2_B$ as a function of $\fnl$ in
Fig.~\ref{figura4} (left panel). Goodness of fit analysis yields
$-300<\fnl<650$ at $68\%$ CL and $-800<\fnl<1050$ at $95\%$, with a
best fit value ($\chi^2_B$ at its minimum) of $\fnl=200$. 
In order to cross check
this result, we estimate a ``frequentist'' confidence interval for
$\fnl$ by sampling the empirical distribution of the $M_J$-based
$\chi^2$, computed for $\fnl=200$. The resulting histogram is displayed
in the right panel of Fig.\ref{figura4}. This analysis shows that we
should expect to constrain $|\fnl|\lesssim 1000$ at $95\%$, thus
suggesting that our limits derived through goodness of fit analysis
are consistent.  One may consider what limits on $\fnl$ would be
derived if we use, in place of MFs, the map's
skweness and kurtosis defined in Sect.~\ref{sec2} as elements of a two
dimensional data vector. We thus repeated our goodness of fit analysis
using these statistics and found weaker limits: $-950<\fnl<1350$ at
$68\%$ CL ($-1400<\fnl<1800$ at $95\%$ CL). Even so, it is quite
remarkable that a crude 1-point pixel statistic degrades the final
sensitivity only by a factor $\sim 2$. Of course, in order to find
``optimal'' constraints one has to implement a more sophisticated
analysis.

The B03 dataset is not expected to be noise dominated at the $6.8'$
resolution employed in the analysis above. To show that this is the
case, we repeated all of our procedures for $13.6'$, finding similar
(though slightly weaker) constraints.

\section{Conclusions}
\label{sec4}
We analysed the B03 $145$~GHz T map in search
of NG signals. We worked in pixel space at $6.8'$ and
$13.6'$ Healpix resolution. We computed the skewness and kurtosis of
the map, as well as its three MFs. We compared these estimates against
a set of simulated Gaussian maps that have been reduced using the same
analysis pipeline as real data, finding no evidence of NG
behavior. To quantify the latter statement, we define goodness of fit
statistics jointly based on all three MFs, showing
that the probability for a Gaussian simulation to have a $\chi^2$
larger than the data is $\sim 67\%$. Assuming a model for primordial
fluctuations that predicts a quadratic perturbation to the
gravitational potential, we set limits on the non linear coupling
parameter $\fnl$ as $-800<\fnl<1050$ at $95\%$ CL ($-300<\fnl<650$ at
$68\%$ CL). These limits may be regarded as complementary
to the constraints set by WMAP in view of the better signal to noise
ratio at high resolution in the B03 field.

\acknowledgements{BOOMERanG was supported by
CIAR, CSA and NERSC in Canada, ASI, University La Sapienza and PNRA in
Italy, NASA and NSF in the US. We thank CASPUR and NERSC/LBL. We acknwoledge the use of HEALPix.}


\begin{thebibliography}{37}
\expandafter\ifx\csname natexlab\endcsname\relax\def\natexlab#1{#1}\fi
\bibitem[Bartolo et al.\ 2004]{bartolo}Bartolo N., Komatsu E., Matarrese S., \& Riotto A., Phys. Rept., 402, 103, 2004
\bibitem[Komatsu and Spergel 2001]{komatsu_spergel} Komatsu E., Spergel D.N., 
Phys.Rev., D63, 063002
\bibitem[Komatsu et al.\ 2003]{komatsu} Komatsu E. et al.,Astrophys.J.Suppl. 148 (2003) 119-134, 2003
\bibitem[Santos et al.\ 2003]{santos} Santos M.G., et al., MNRAS, 341, 623, 2003
\bibitem[Wu et al.\ 2001]{wu} Wu J.H.P., et al., Phys.Rev.Lett. 87 (2001) 251303
\bibitem[Smith et al.\ 2004]{vsa} Smith S., et al., MNRAS, 352, 887, 2004
\bibitem[Curto et al. \ 2006]{archeops} Curto A., et al., astro-ph/0612148, 
submitted to A\& A, 2006
\bibitem[Polenta et al.\ 2002]{polenta} Polenta G., et. al., 2002, ApJ., 572 L27
\bibitem[De Troia et al.\ 2003]{detroia} De Troia G., et. al., 2003 MNRAS, 343, 284
\bibitem[Spergel et al.\ 2006]{spergel} Spergel D.N., et al., astro-ph/0603449,
ApJ, in press
\bibitem[Creminelli et al.\ 2007]{creminelli} Creminelli, P., et al.,  JCAP, 3, 5, 2007 
\bibitem[Copi et al.\ 2004]{copi}  Copi C.J.,Huterer D.,Starkman G.D.,Phys.Rev.,D70, 043515, 2004 
\bibitem[Vielva et al.\ 2004]{vielva} Vielva P., et al., ApJ, 609, 22-34, 2004
\bibitem[Cruz et al.\ 2006]{cruz} Cruz M., et al., MNRAS, 369, 57-67, 2006
\bibitem[Liguori et al.\ 2003]{liguori} Liguori M., Matarrese S.\ and Moscardini L., ApJ., 597, 57, 2003
\bibitem[Masi et al.\ 2005]{masi} Masi S. et al., astro-ph/0507509, submitted to A\&A, 2005
\bibitem[Jones et al.\ 2006]{jones} Jones W.C., et. al., Astrophys.J. 647, 823, 2006 
\bibitem[Montroy et al.\ 2006]{montroy} Montroy T.E., et al., Astrophys.J. 647, 813, 2006 
\bibitem[Piacentini et al.\ 2006]{piacentini} Piacentini F., et al., Astrophys.J. 647, 833, 2006 
\bibitem[MacTavish et al.\ 2006]{mactavish} MacTavish C.J.,  et al., Astrophys.J. 647, 799, 2006
\bibitem[Natoli et al.\ 2001]{natoli} Natoli P., et al.\ , A\&A, 371, 346, 2001  
\bibitem[De Gasperis et al.\ 2005]{degasperis} de Gasperis G., et al., A\&A, 436, 1159, 2005 
\bibitem[Gott et al.\ 1990]{mink_f} Gott J.R., et al., ApJ., 352, 1, 1990
\bibitem[Cabella et al.\ 2004]{cabella} Cabella P., et al., Phys.Rev. D69, 
063007, 2004
\bibitem[Schmalzing and G\'orski 1998]{schmalzing} Schmalzing J.\ and G\'orski K.M.\ , MNRAS, 297, 355, 1998
\bibitem[Winitzki and Kosowsky 1998]{winitzki} Winitzki S.\ and Kosowsky A.\ , New Astron.\ 3, 75, 1998
\bibitem[G\'orski et al.\ 2005]{healpix} G\'orski K. M., et al., ApJ, 622, 759, 2005 

\end{thebibliography}
\end{document}